\documentclass[english,conference]{IEEEtran}
\usepackage[T1]{fontenc}
\usepackage[latin9]{inputenc}
\usepackage{amsmath}
\usepackage{graphicx}

\makeatletter

\providecommand{\tabularnewline}{\\}

\makeatother

\usepackage{babel}
\begin{document}
\title{Physics Informed Neural Network using Finite Difference Method}
\author{Kart Leong Lim, Rahul Dutta, Mihai Rotaru}
\maketitle
\begin{abstract}
In recent engineering applications using deep learning, physics-informed
neural network (PINN) is a new development as it can exploit the underlying
physics of engineering systems. The novelty of PINN lies in the use
of partial differential equations (PDE) for the loss function. Most
PINNs are implemented using automatic differentiation (AD) for training
the PDE loss functions. A lesser well-known study is the use of finite
difference method (FDM) as an alternative. Unlike an AD based PINN,
an immediate benefit of using a FDM based PINN is low implementation
cost. In this paper, we propose the use of finite difference method
for estimating the PDE loss functions in PINN. Our work is inspired
by computational analysis in electromagnetic systems that traditionally
solve Laplace's equation using successive over-relaxation. In the
case of Laplace's equation, our PINN approach can be seen as taking
the Laplacian filter response of the neural network output as the
loss function. Thus, the implementation of PINN can be very simple.
In our experiments, we tested PINN on Laplace's equation and Burger's
equation. We showed that using FDM, PINN consistently outperforms
non-PINN based deep learning. When comparing to AD based PINNs, we
showed that our method is faster to compute as well as on par in terms
of error reduction.
\end{abstract}

\section{Introduction}

Recent advancement in multiphysics modeling combines deep learning
with partial differential equations (PDEs), through the use of modern
deep learning tools. The physics-informed neural network (PINN) \cite{raissi2019physics}
play such a role by constraining the data-driven neural network with
physics laws. A key aspect of PINN is the use of automatic differentiation
(AD) which compute partial derivatives of the neural network, allowing
PDEs to seamlessly backpropagate through neural network \cite{raissi2019physics,Yang2020PhysicsInformedGA,yang2019adversarial,zhong2022pi,kim2020dpm,chen2021physics}.
This physics-awareness \cite{patel2022thermodynamically} which transits
from physics law to weight update allows neural network to better
generalize real-world problem. An awareness that is unlikely to be
picked up by past data-centric deep learning approaches such as ResNet
\cite{he2016deep}. Till date, PINNs are seen in fluid mechanics \cite{raissi2019physics},
heat transfer \cite{hennigh2021nvidia}, contaminant transport \cite{praditia2021finite},
shock hydrodynamics and electromagnetic field \cite{fang2019deep}.

Despite the success of using AD in PINN, some researchers \cite{9403414,chiu2021canpinn}
have contemplated AD as a black-box solver and brought up several
concerns. Problems arising from AD based PINN includes: 
\begin{quote}
i) Having lower accuracies when smaller collocation points are used
\cite{chiu2021canpinn} and longer computational time.

ii) Seldom shown theoretically to converge \cite{9403414}. 
\end{quote}
To the best of our knowledge, only a handful of numerical differentiation
(ND) themed PINN exists. They include: a) PINN using finite volume
method \cite{9403414}, b) PINN which combines automatic difference
with finite difference \cite{chiu2021canpinn}, c) control volume
PINN \cite{patel2022thermodynamically} and the finite volume PINN
\cite{praditia2021finite}. To distinguish between these PINNs, we
group the former \cite{chiu2021canpinn,9403414} as PINNs which discard
automatic differentiation from the training of PINN entirely while
the latter \cite{praditia2021finite,patel2022thermodynamically} are
PINNs that incorporate NDs into the loss functions but will require
automatic differentiation to work.

Our task is mainly targeted at exploring numerical differentiation
as an alternative to automatic differentiation for the PINN loss function
(i.e. following the works of \cite{chiu2021canpinn,9403414}). We
quantify this task assessment by measuring the efficiency of PINNs
under fewer collocation points and comparing the computational cost.
Our approach is inspired by electromagnetic analysis that uses finite
difference method (FDM) to approximate PDEs. We first disjoint the
PDE solver from the loss function of PINN, simplifying our task. The
PDE solver can be as simple as averaging the neighborhood of a node.
In the case of Laplace's equation, our PDE solver using FDM mimics
the residual node response of electromagnetic system and resembles
the Laplacian filter in image processing. Using chain rule, the PINN
loss function can be re-assembled back and can be trained completely
without requiring AD. We called our method as the FDM based PINN (FDM-PINN). 

In our experiment, we applied FDM-PINN to the rectangular conducting
trough problem which uses Laplace's equation and the classical case
of PINN using Burger's equation. We compared our approach to both
the AD based PINN and a regular neural network or NN. Our results
showed that FDM-PINN outperforms NN in terms of lower error. Also,
FDM-PINN does not require a GPU to compute, yet is on par or outperforms
the AD based PINN at low collocation points conditions and is computational
faster.

To our best knowledge, the only relevant work to FDM-PINN is can-PINN
\cite{chiu2021canpinn}. The main difference being can-PINN is a hybrid
solution that first requires automatic differentiation to compute
the PDE loss i.e. $\delta u/\delta x$ (where $u$ is the neural network
output) before applying numerical differentiation on $\delta u/\delta x$
to augment PINN. Furthermore, can-PINN is only tested on first-order
PDE problems. In our case, we are completely automatic differentiation
free when training the PDE loss of PINN. This is possible by introducing
a PDE solver \cite{nagel2011solving} to the PDE loss of PINN as discussed
in Section III.C.

\section{Background on PINN}

\begin{figure}
\begin{centering}
\includegraphics[scale=0.55]{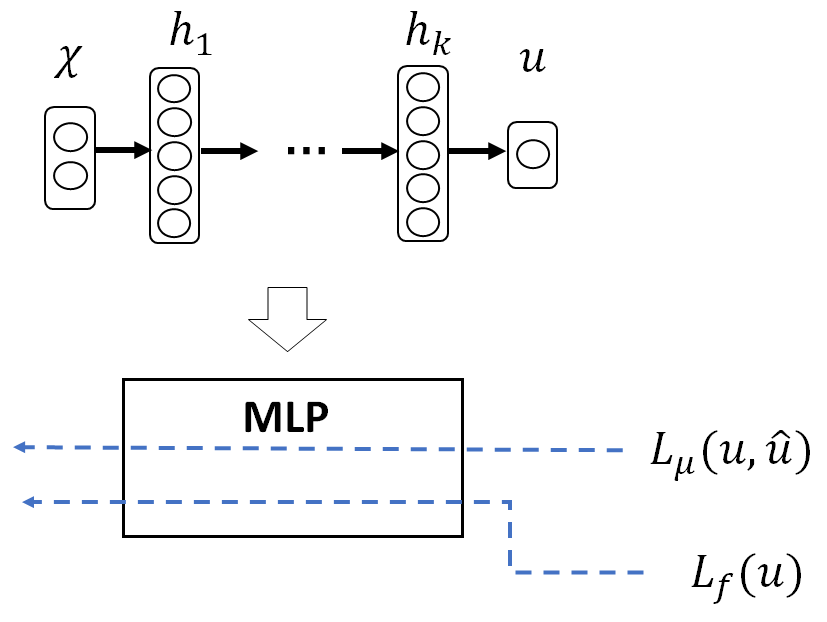}
\par\end{centering}
\caption{PINN is trained using mean square error loss $L_{\mu}$ (data-driven)
and physics-informed loss $L_{f}$ (data-free). The architecture of
PINN uses a MLP.}
\end{figure}

In recent engineering applications using deep learning, physics-informed
neural network (PINN) stood out as it can exploit the underlying physics
of engineering systems. The novelty of PINN lies in the use of partial
differential equations (PDE) as regularizer loss.

Unlike deep learning \cite{krizhevsky2012imagenet} which requires
training label for training the architecture, the regularizer in PINN
does not require training labels. Instead, we train the regularizer
loss in PINN similarly to a VAE's regularizer loss. Examples of neural
networks using regularizer loss includes Autoencoder \cite{hinton2006reducing},
Variational autoencoder (VAE) \cite{kingma2014stochastic} and Generative
adversarial net (GAN) \cite{goodfellow2014generative}. A basic PINN
architecture uses a standard multilayer perceptron (MLP). While VAE
formulates the latent space of an Autoencoder using probability distributions,
PINN formulates the output space of MLP using PDEs. We illustrate
the loss functions of PINN in Fig. 1. 

The PINN objective \cite{raissi2019physics} seeks to minimize the
losses incurred by i) mean-square error loss $L_{\mu}$ and ii) physics-informed
loss $L_{f}$ as follows where $\lambda$ is the tuning parameter.

\begin{equation}
L_{PINN}=L_{\mu}+\lambda L_{f}
\end{equation}

\subsection{Mean square error loss}

For Burger's equation, we refer to the input variables $t$ and $x$
as the time and sinusoidal axis in Fig. 4. The training dataset is
defined as $\chi=\left\{ t_{m},x_{m}\right\} _{m=1}^{M}$ and $\hat{u}=\left\{ \hat{u}_{m}\right\} _{m=1}^{M}$
for the input and output respectively for sample size $M$. For some
multiphysics model e.g. Burger's equation, the training dataset can
be obtained by generating dataset from initial condition and boundary
conditions. For the $b^{th}$ initial or boundary condition, the output
function $g_{b}$ and its inputs $t_{b},\;x_{b}$ are given as

\begin{equation}
\hat{u}_{b}=g_{b}\mid_{t_{b},\;x_{b}}
\end{equation}

Whereas for Laplace's equation, we refer to the input variables as
$\chi=\left\{ x_{m},y_{m}\right\} _{m=1}^{M}$ and output variable
$\hat{V}=\left\{ \hat{V}_{m}\right\} _{m=1}^{M}$ for the 2D electromagnetic
field in Fig. 3.

$\;$

The supervised learning part of PINN minimizes the Euclidean distance
between ground truth $\hat{V}$ and prediction $V$ using mean square
error (MSE) loss which is defined as (vice versa for $\hat{u}$ and
$u$)

\begin{equation}
L_{\mu}=\frac{1}{M}\sum_{m=1}^{M}\left|\hat{V}-V\right|^{2}
\end{equation}

\subsection{Physics-informed loss}

For the unsupervised learning part of PINN, the physics-informed loss
is a regularization that enforces PDE constraints on PINN by observing
the prediction $u$. The PDE, $f(u)$ captures the known underlying
physics of an engineering system as studied by domain experts. The
physics-informed loss is defined as

\begin{equation}
L_{f}=\frac{1}{M}\sum_{m=1}^{M}\left|f(u)\right|^{2}
\end{equation}

\begin{figure}
\begin{centering}
\includegraphics[scale=0.7]{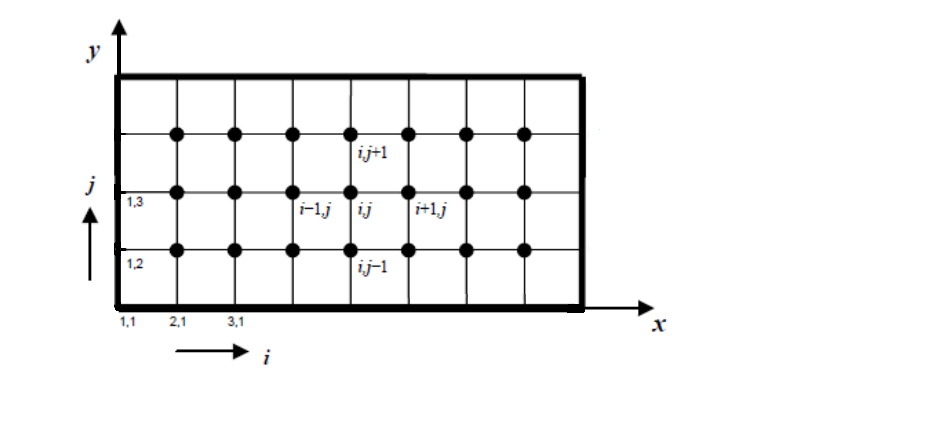}
\par\end{centering}
\caption{Approximating continuous electromagnetic field $V_{(x,y)}$, using
a discrete 2D uniform grid of nodes $\varGamma_{(i,j)}$.}
\end{figure}

\section{Proposed method }

In this section, we mainly focus on the estimation of PDEs in physics-informed
loss using numerical differentiation to reduce the complexity of training.
In fact, numerical differentiation has been the de facto for multiphysics
modeling for the past few decades and can be categorized into i) finite
element method (FEM), ii) finite volume method (FVM), iii) finite
difference method (FDM) and iv) boundary element method (BEM). Inspired
by electromagnetic analysis \cite{nagel2011solving}, we propose to
estimation PDEs using finite difference method (FDM). As mentioned
earlier, our novelty is not about which numerical method is better
e.g. FEM, FDM or FVM. Instead, our novelty lies in proposing the use
of numerical differentiation for PINN over automatic differentiation
\cite{baydin2018automatic}. 

We demonstrate the proposed method on classical PDEs such as Laplace's
and Burger's equations. The former is applied to electromagnetic systems
and the latter can be found in fluid mechanics systems.

\subsection{Laplace's Equation}

In electromagnetic systems, the electromagnetic field is governed
by Laplace's equation (in rectangular coordinates) in eqn (6). The
estimation to Laplace's equation can be performed using a finite difference
method in electromagnetic analysis known as the successive over-relaxation
(SOR) algorithm \cite{nagel2011solving}. The standard procedure for
SOR is summarized as follows: 
\begin{verse}
i) Discretize a 2D uniform grid of nodes for representing the electromagnetic
field $V$

ii) Slide the window computation of the residual $R$ across each
rows and columns. 
\end{verse}
Specifically, each iteration of SOR computes the residual $R(i,j)$
per node, before applying an incremental update $V_{new}(i,j)$ as
follows

\begin{equation}
\begin{array}{c}
R(i,j)=\frac{1}{4}\left[V_{(i-1,j)}+V_{(i,j-1)}\right.\\
\left.+V_{(i,J+1)}+V_{(i+1,j)}\right]\\
\\
V_{new}(i,j)=V_{old}(i,j)+R(i,j)\\
\\
\end{array}
\end{equation}

Typically, SOR takes $T$ numbers of iterations to converge. The number
of $T$ increases as the resolution of $V(i,j)$ increases.

\begin{figure}
\begin{centering}
\includegraphics[scale=0.3]{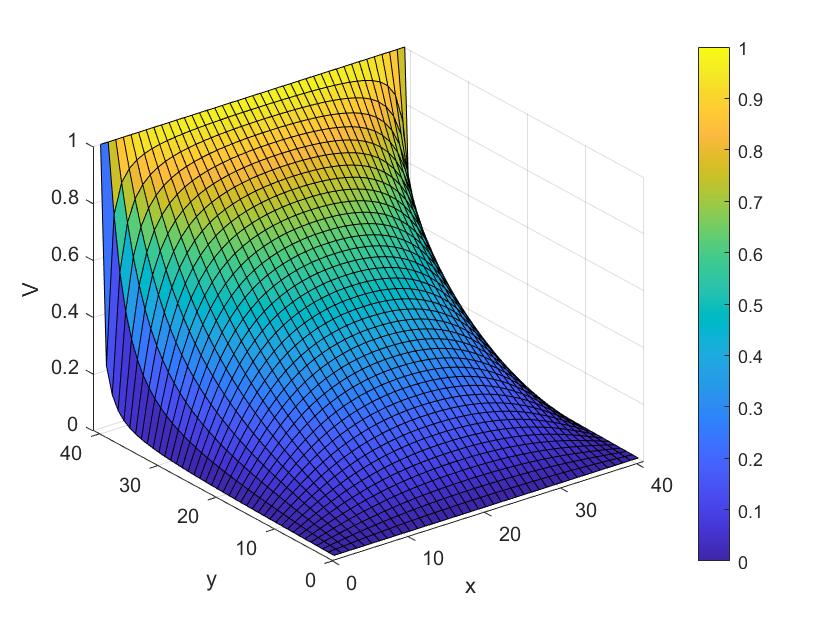}
\par\end{centering}
\caption{Plot of Laplace's equation (rectangular conducting trough) generated
using SOR.}
\end{figure}

The rectangular conducting trough is an example of solving Laplace's
equation (with Dirichlet boundary conditions in Table II) found in
electromagnetic system \cite{sadiku2015analytical}. The PDE (boundary
conditions in Table II) is given as follows

\begin{equation}
f(V)=\frac{d^{2}V}{dx^{2}}+\frac{d^{2}V}{dy^{2}}
\end{equation}

When modeling $f(V)$ as a physics-informed loss in Fig 2, it is non-trival
to define a closed form equation for backpropagation. Inspired by
SOR where $f(V)$ is approximated using a finite difference method
(FDM) \cite{nagel2011solving}, we sought to apply FDM to physics-informed
loss. First, we discretize $f(V)$ over a grid of uniformly distributed
nodes as shown in Fig. 1. Where $h$ is the distance between each
node. Next, we approximate $f(V)$ in eqn (6) using FDM (in Table
I) which becomes the following expression

\begin{equation}
\varGamma_{(i,j)}=\frac{V_{(i-1,j)}+V_{(i+1,j)}+V_{(i,j-1)}+V_{(i,j+1)}-4V_{(i,j)}}{h^{2}}
\end{equation}

$\;$

We refers to $\varGamma_{(i,j)}$ as the FDM of $f(V)$ for a single
node at $(i,j)$. We also note that eqn (7) is identical to SOR in
eqn (5). Coincidentally, $\varGamma_{(i,j)}$ resembles the Laplacian
filter for edge detection in image processing.

\subsection{Burger's Equation}

\begin{figure}
\begin{centering}
\includegraphics[scale=0.32]{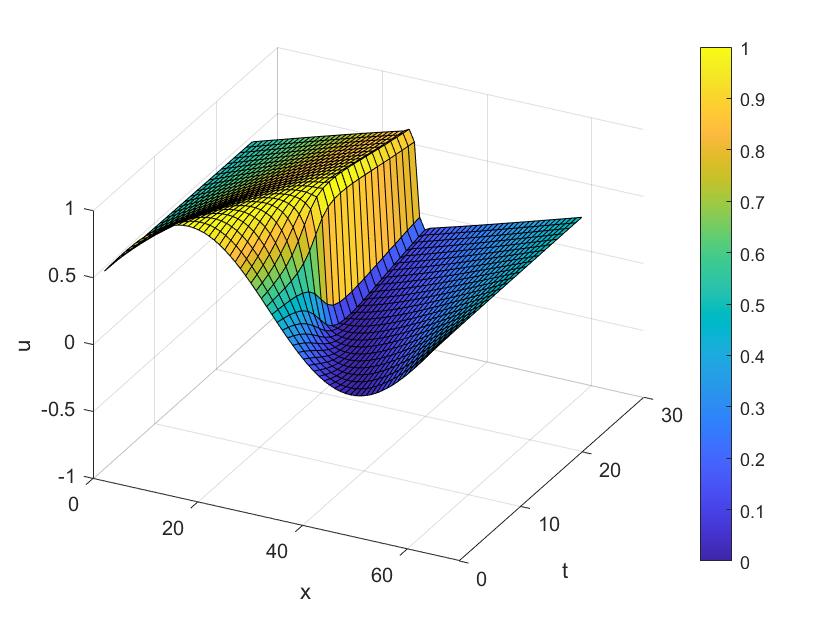}
\par\end{centering}
\caption{Plot of Burger's equation generated using ``burgers\_shock.mat''
\cite{raissi2019physics}.}
\end{figure}

We propose on how FDM can be applied to PINN for solving Burger's
equation. The PDE of Burger's equation (boundary conditions in Table
II) is defined as \cite{raissi2019physics}

\begin{equation}
f(u)=u_{t}+uu_{x}-\left(\frac{0.01}{\pi}\right)u_{xx}
\end{equation}

$\;$

Since $f(u)$ is intractable, most authors use automatic differentiation
library e.g. ``tf.gradients'' or ``dlgradient'' in Table VI to
handle the PDE computations. Instead, we propose to approximate $f(u)$
using FDM. First we discretize the continuous time model $u(x,t)$
by defining a 2D grid of uniformly spaced nodes over it. Next, we
use FDM in Table I to approximate the first-order and second-order
partial derivatives for the PDE. Thus, we arrive at the following
equation

\pagebreak{}

\[
\Gamma_{(i,j)}=\frac{-3u_{(i,j)}+4u_{(i,j+1)}-u_{(i,j+2)}}{2h}
\]

\[
+u_{(i,j)}\left[\frac{-3u_{(i,j)}+4u_{(i+1,j)}-u_{(i+2,j)}}{2h}\right]
\]

\begin{equation}
-\left(0.01/\pi\right)\frac{u_{(i-1,j)}-2u_{(i,j)}+u_{(i+1,j)}}{h^{2}}
\end{equation}

$\;$

\subsection{PINN Weight Update}

Once we have computed $\Gamma_{(i,j)}$, we can easily train PINN
for both Burger's and Laplace's equation case. We seek to compute
updates for the MLP weight, $w_{j,j+1}$ using Stochastic Gradient
Ascent (SGA) shown below and illustrated in Fig. 1 

\begin{equation}
\triangle w_{j,j+1}=\gamma_{_{\mu}}\left(\nabla L{}_{\mu}\right)+\lambda\gamma_{_{f}}\left(\nabla L{}_{f}\right)
\end{equation}

In eqn (10), the weight update of the hidden layers is defined as
$\triangle w_{j,j+1}$ for $j=1...k$ number of hidden layers and
learning rates $\gamma_{_{\mu}},\gamma_{_{f}}$. $\nabla L{}_{\mu}$
refers to the gradient of the MSE loss and closed form backpropagation
is available \cite{hinton2006reducing}. On the other hand for the
gradient of the PINN loss, $\nabla L{}_{f}$ is non-trival to compute.
Thus, we propose to breakdown $\nabla L{}_{f}$ using chainrule in
eqn (11) to include the FDM term $\varGamma$, as previously introduced
in eqn (7) and (9) for Laplace and Burger equation respectively

\begin{equation}
\nabla L{}_{f}=\frac{\delta L_{f}}{\delta\varGamma}\times\frac{\delta\varGamma}{\delta V}
\end{equation}

In eqn (11), $\delta L_{f}/\delta\varGamma$ is a differentiable term
as $L_{f}=\frac{1}{M}\sum_{m=1}^{M}\left|\varGamma\right|^{2}$ is
a L2 regularizer loss. However, $\delta\varGamma/\delta V$ represents
the gradient of a local surface bounded by $V_{(i,j)}$ and its neighboring
nodes. For tractability, we further assume $\delta\varGamma/\delta V$
is a constant. 

\section{Experiments}

In Table II, we refer to the PINN setup for modeling Laplace's and
Burger's equation. The setup describes parameters for the continuous
inputs and outputs, training dataset, initial / boundary conditions
and their respective sample sizes $\left(N_{ini},N_{bc}\right)$,
collocation points $\left(N_{f}\right)$, loss functions, no. of neurons/
hidden layers and SGA parameters. We also restrict to lower collocation
points to demonstrate the advantages of ND based PINN.

We perform the modeling tasks using:
\begin{verse}
a) NN: training PINN using MSE loss $L_{\mu}$ only.

b) FDM-PINN (proposed): training PINN using $L_{PINN}$ where physics-informed
loss $L_{f}$ is using finite difference method described in Section
III. 

c) AD-PINN: training PINN using $L_{PINN}$ where $L_{f}$ is computed
by automatic differentiation found in deep learning libraries e.g.
Table VI.
\end{verse}
\begin{table}
\centering{}\caption{Approximating partial differential equations using finite difference
method.}
\begin{tabular}{|c|c|}
\hline 
First order PDE & $\begin{array}{c}
\\
\frac{du}{dx}=\frac{-3u_{(i,j)}+4u_{(i,j+1)}-u_{(i,j+2)}}{2h}\\
\\
\end{array}$\tabularnewline
\hline 
Second order PDE & $\begin{array}{c}
\\
\frac{d^{2}u}{dx^{2}}=\frac{u_{(i-1,j)}-2u_{(i,j)}+u_{(i+1,j)}}{h^{2}}\\
\\
\end{array}$\tabularnewline
\hline 
\end{tabular}
\end{table}

\begin{table*}
\centering{}\caption{Experimental Setup}
\begin{tabular}{|c|c|c|}
\hline 
 & Laplace's equation, $V(x,y)$ & Burger's equation, $u(x,t)$\tabularnewline
\hline 
$\begin{array}{c}
Continuous\\
inputs\;and\;output
\end{array}$  & $\begin{array}{c}
\\
x\in[0,1],\;y\in[0,1],\;V\in[0,1]\\
\\
\end{array}$ & $\begin{array}{c}
\\
x\in[-1,1],\;t\in[0,1],\;u\in[-1.0131,1.0131]\\
\\
\end{array}$\tabularnewline
\hline 
$\begin{array}{c}
Ground\\
truth
\end{array}$ & generated by SOR in eqn (6) & ``burgers\_shock.mat'' \cite{raissi2019physics}\tabularnewline
\hline 
$\begin{array}{c}
\#\;Nodes\end{array}$ & $N_{i}\times N_{j}=41\times41$ & $N_{i}\times N_{j}=64\times25$\tabularnewline
\hline 
$\begin{array}{c}
Initial/Boundary\\
conditions
\end{array}$ & $\begin{array}{c}
\\
\hat{V}_{1}=0\mid{}_{x_{1}=0,\;y_{1}=y}\\
\hat{V}_{2}=0\mid{}_{x_{2}=x,\;y_{2}=0}\\
\hat{V}_{3}=0\mid{}_{x_{3}=1,\;y_{3}=y}\\
\hat{V}_{4}=1\mid{}_{x_{4}=x,\;y_{4}=1}\\
\\
\end{array}$ & $\begin{array}{c}
\\
\hat{u}_{1}=-\sin(\pi x)\mid{}_{x_{1}=x,\;t_{1}=0}\\
\hat{u}_{2}=0\mid{}_{x_{2}=-1,\;t_{2}=t}\\
\hat{u}_{3}=0\mid{}_{x_{3}=1,\;t_{3}=t}\\
\\
\end{array}$\tabularnewline
\hline 
$Initial\;condition\;samples,N_{ini}$ & $20,30,50,100,1000$ & $10,30,50,100$\tabularnewline
\hline 
$Boundary\;condition\;samples,N_{bc}$ & 20 & $10,30,50,100$\tabularnewline
\hline 
$Collocation\;points,N_{f}$ & 1681 & 1600\tabularnewline
\hline 
$\begin{array}{c}
Loss\\
functions\\
(\lambda=0.7)
\end{array}$ & \multicolumn{2}{c|}{$\begin{array}{c}
\\
L_{PINN}=L_{\mu}+\lambda L_{f}\\
\\
\end{array}$}\tabularnewline
\hline 
$Network\;architecture$ & $2-50-50-50-1$ & $2-30-30-30-1$\tabularnewline
\hline 
$Activation\;function$ & \multicolumn{2}{c|}{$tanh$}\tabularnewline
\hline 
$PINN\;learner$ & \multicolumn{2}{c|}{SGA with momentum}\tabularnewline
\hline 
$Learning\;rates$ & \multicolumn{2}{c|}{$\gamma{}_{\mu}=1e{}^{-1}$, $\gamma{}_{f}=1e^{-3}\sim1e{}^{-5}$}\tabularnewline
\hline 
$Iterations\;ran$ & \multicolumn{2}{c|}{10,000}\tabularnewline
\hline 
$minibatch\;size\;for\;L_{f}$  & 32 & 8\tabularnewline
\hline 
$minibatch\;size\;for\;L_{\mu}$  & 4 & 8\tabularnewline
\hline 
$Evaluation\;metric$ & $\begin{array}{c}
\begin{array}{c}
\\
L2=\sqrt{\sum_{j=1}^{N_{j}}\sum_{i=1}^{N_{i}}\left(\hat{V_{\left(i,j\right)}}-V_{\left(i,j\right)}\right)^{2}}\\
\\
\end{array}\end{array}$ & $\begin{array}{c}
\\
L2=\frac{1}{N_{j}}\sum_{j=1}^{N_{j}}\sqrt{\sum_{i=1}^{N_{i}}\left(u_{\left(i,j\right)}-\hat{u_{\left(i,j\right)}}\right)^{2}}\\
\\
\end{array}$\tabularnewline
\hline 
\end{tabular}
\end{table*}

\subsection{Burger's equation}

The main difference between Burger's and Laplace's equation is the
implementation of $\varGamma_{(i,j)}$ in eqn (13) and (7). The minor
differences are the parameters such as the network architecture, and
grid size, $N_{ini}$ and etc in Table II. 

The ground truth (Fig. 4) for Burger's equation is available online.
We use a 64x25 grid for the ground truth. Our experiment is follows
some of the setup in \cite{raissi2019physics} where possible. We
tabulate the average of 10 runs results for NN, AD-PINN and FDM-PINN
and refer to the L2 error in Table II. Instead of computing L2 at
every node, we use L2 to measure the full sinusoidal response between
ground truth and PINN prediction at each time $t$. We then take the
average of $u(x,t)$ for all $t$ as the error. In Table III, all
methods suffer from larger error when training samples $N_{ini}$
and $N_{bc}$ are too low at 10 samples each. Both NN and FDM-PINN
are implemented using eqn (11). The main difference between NN and
FDM-PINN is the additional $L_{f}$ term in eqn (9). Thus, we can
find out how much $L_{f}$ reduces the error. We observe that in fact
FDM-PINN outperforms NN significantly. Both FDM-PINN and AD-PINN uses
the same architecture setup and iterations ran and training samples
counts. However, AD-PINN uses automatic differentiation for solving
$L_{f}$, $L_{\mu}$ and the MLP. From the results, FDM-PINN is on
par or better than AD-PINN for all training conditions. In terms of
code execution time (Table V), FDM-PINN takes approximately 5X longer
than NN to compute. Whereas, AD-PINN takes 5X longer than FDM-PINN,
despite that AD-PINN uses a laptop based GPU whereas FDM-PINN and
NN are computed using CPU. We suspect this could be due to the multiple
AD operations required to run Burger's equation as compared to the
proposed FDM approach. 

\begin{table}
\centering{}\caption{MSE on Burger's equation}
\begin{tabular}{|c|c|c|c|c|}
\hline 
$N_{ini}$ & 10 & 30 & 50 & 100\tabularnewline
$N_{bc}$ & 10 & 30 & 50 & 100\tabularnewline
$N_{f}$ & 1600 & 1600 & 1600 & 1600\tabularnewline
\hline 
NN & 2.0133 & 1.8493 & 1.6720 & 1.7107\tabularnewline
\hline 
AD-PINN & 1.6450 & \textbf{1.3291} & \textbf{1.3293} & 1.4767\tabularnewline
\hline 
FDM-PINN & \textbf{1.5629} & 1.3302 & 1.3426 & \textbf{1.4153}\tabularnewline
\hline 
\end{tabular}
\end{table}

\subsection{Laplace's equation}

\begin{table}
\centering{}\caption{MSE on Laplace's equation}
\begin{tabular}{|c|c|c|c|c|c|}
\hline 
$N_{ini}$ & 20 & 30 & 50 & 100 & 1000\tabularnewline
$N_{bc}$ & 20 & 20 & 20 & 20 & 20\tabularnewline
$N_{f}$ & 1681 & 1681 & 1681 & 1681 & 1681\tabularnewline
\hline 
NN & 5.1164 & 4.8198 & 4.3277 & 4.0809 & 4.1246\tabularnewline
\hline 
AD-PINN & 5.1946 & 4.8686 & 4.9758 & 4.5999 & 4.1131\tabularnewline
\hline 
FDM-PINN & \textbf{4.5782} & \textbf{3.6802} & \textbf{3.9836} & \textbf{3.4536} & \textbf{3.6162}\tabularnewline
\hline 
\end{tabular}
\end{table}

We use SOR to generate a 41x41 ground truth for the ``rectangular
conducting trough'' in Fig. 3. The trough problem does not have a
time variable like Burger's equation. Instead, we have a 2D plane
with surrounding boundary conditions \cite{sadiku2015analytical}.
Thus, we refer to $N_{ini}$ instead as random subsamples from the
ground truth. $N_{bc}$ refers to random samples drawn from the 4
boundary conditions. Comparison of NN and PINN are made for fixed
boundary condition sample size $N_{bc}$ and varying initial sample
size $N_{ini}$ in Table III. We measure the L2 error between ground
truth and PINN output for the entire grid. We tabulate the average
of 10 runs results for the NN, FDM-PINN and AD-PINN in Table IV. Firstly,
we observe that a larger $N_{ini}$ reduces the error for all methods.
Secondly, we observe that NN and AD-PINN are quite similar in performance.
However, AD-PINN suffers from poorer result between $N_{ini}$=50
and $N_{ini}$=100. Overall, FDM-PINN outperforms both NN and AD-PINN
significantly. The larger minibatch size for $L_{f}$ as compared
to $L_{\mu}$ in Table II is found to empirically give a better result
for FDM-PINN. However, the same setting did not appear to benefit
AD-PINN. Also, increasing the minibatch size for $L_{\mu}$ beyond
8 did not change the result much for both NN, AD-PINN and FDM-PINN.
In terms of execution time in Table V, FDM-PINN is slightly faster
to compute than AD-PINN (with GPU). Due to the increase in hidden
layers' neuron and minibatch size for $L_{f}$, FDM-PINN takes longer
to compute Laplace's equation than previously for Burger's equation
in Table V. 

\begin{table}
\centering{}\caption{Execution time (per iteration)}
\begin{tabular}{|c|c|c|}
\hline 
 & Burger's Equation & Laplace's Equation\tabularnewline
\hline 
NN & 0.0023 sec & 0.0028 sec\tabularnewline
\hline 
AD-PINN & 0.0478 sec & 0.0653 sec\tabularnewline
\hline 
FDM-PINN & 0.0097 sec & 0.0609 sec\tabularnewline
\hline 
\end{tabular}
\end{table}

\begin{table}
\centering{}\caption{Examples of using automatic differentiation for Burger's equation}
\begin{tabular}{|c|c|c|}
\hline 
 & Matlab & Tensorflow\tabularnewline
\hline 
\hline 
$u=$ & $model(parameters,x,t)$ & $model(self,x,t,parameters)$\tabularnewline
\hline 
$u_{x}=$ & $dlgradient(u,x)$ & $tf.gradients(u,x)[0]$\tabularnewline
\hline 
$u_{t}=$ & $dlgradient(u,t)$ & $tf.gradients(u,t)[0]$\tabularnewline
\hline 
$u_{xx}=$ & $dlgradient(u_{x},x)$ & $tf.gradients(u_{x},x)[0]$\tabularnewline
\hline 
\end{tabular}
\end{table}

\section{Conclusion and future work }

Automatic differentiation is a modern deep learning blackbox that
enables PINNs to constraint the learning of neural networks with physics
laws. On the other hand, numerical differentiation has always been
the de facto PDE solver in multiphysics simulations for the past few
decades. For instance, in electromagnetic analysis, the successive
over-relaxation algorithm is a well known example of numerical differentiation
that approximates the PDE of Laplace's equation. 

In our work, we propose using numerical differentiation over automatic
differentiation when training PINN. In our approach, we decouple the
PDE solver from the PINN loss function such that it does not require
automatic differentiation to be solved. Instead, we use numerical
differentiation for the PDEs which in the Laplace's equation case
is as simple as taking the average on the neighborhood of a node.
This method can also be straightforward extended to the Burger's equation.
We tested our proposed method (FDM-PINN) on both datasets of Laplace's
and Burger's equation. We made comparison with the automatic differentiation
based PINN. Our results under the condition of low collocation points
showed that FDM-PINN is faster to compute while at least on par or
better than automatic differentiation based PINN. However, one of
main drawback FDM-PINN currently face is the need to define its collocation
points on a grid. Secondly, we currently do not have a solution to
compute the gradient of node surface i.e. $\delta\varGamma/\delta u$. 

In future, we may consider a Markov-random field approach for the
PDE solver since the PINN output is usually homogeneous. Another direction
is to extend the PINN architecture to 2D such as the convolutional
neural net.

\bibliographystyle{IEEEtran}
\addcontentsline{toc}{section}{\refname}\bibliography{allmyref}

\end{document}